# Regularization Methods for Generative Adversarial Networks: An Overview of Recent Studies

Minhyeok Lee[1,2] & Junhee Seok[1]

[1] Electrical Engineering, Korea University, Seoul, Republic of Korea
[2] Research Institute for Information and Communication Technology, Korea University, Seoul, Republic of Korea
[*suam6409, jseok14*]@korea.ac.kr

## Abstract

Despite its short history, Generative Adversarial Network (GAN) has been extensively studied and used for various tasks, including its original purpose, i.e., synthetic sample generation. However, applying GAN to different data types with diverse neural network architectures has been hindered by its limitation in training, where the model easily diverges. Such a notorious training of GANs is well known and has been addressed in numerous studies. Consequently, in order to make the training of GAN stable, numerous regularization methods have been proposed in recent years. This paper reviews the regularization methods that have been recently introduced, most of which have been published in the last three years. Specifically, we focus on general methods that can be commonly used regardless of neural network architectures. To explore the latest research trends in the regularization for GANs, the methods are classified into several groups by their operation principles, and the differences between the methods are analyzed. Furthermore, to provide practical knowledge of using these methods, we investigate popular methods that have been frequently employed in state-of-the-art GANs. In addition, we discuss the limitations in existing methods and propose future research directions.

## 1. Introduction

Generative Adversarial Network (GAN) is one of the most rapidly developed deep learning methods in recent years. While the initial model of GAN has lately been introduced, compared to other conventional deep learning methods, such as Convolutional Neural Network (CNN) and Long Short-Term Memory (LSTM), GAN is now extensively used in various practical applications, including image synthesis [5,33,34,79,82], image inpainting [78], video generation [67], data augmentation [6,44], and style transfer [3,12,54].

Distinct from the other conventional deep learning methods, GAN is a generative model of which objective is to learn distributions instead of data points as targets. Therefore, the output of GAN models is generally distributions as well. By using the Monte Carlo method over the distributions, GAN can be used for synthetic sample generation since the data point extracted by the Monte Carlo method corresponds to a synthetic sample that mimics features of real samples.

However, it is well known that the training of GAN is challenging compared to that of other deep learning methods since the training is commonly unstable, and weight parameters of GAN easily diverge due to its adversarial training process [21]. These incorrectly trained GANs generally produce identical samples regardless of input noises; such a failure of training is called mode collapse problem [77].

Thus, it is extremely critical to stabilize the training of GANs in order to induce GANs to learn certain distributions precisely. Since the convergence of GAN has been proved under the condition that the discriminator in GANs is optimal [20], in recent years, a large number of studies have proposed regularization methods that aim at making the discriminator stable. Also, it has been demonstrated that Lipschitz continuity of the discriminator with regard to sample spaces is a key condition for a stable

GAN training [2]; consequently, numerous methods to enforce the discriminator to satisfy such a condition have been proposed.

In this paper, we review the recent regularization methods for the training of GANs. Specifically, the most recent methods are investigated; then, we classify the regularization methods into several groups according to their operation principles, in order to explore research trends in the regularization methods. Furthermore, popular methods that are frequently employed in state-of-the-art GAN models are investigated.

While several studies have been conducted to review GAN models, they have commonly focused on different architectures of GAN models [24,52], GAN losses [70], or practical applications using GANs [9,76]. Instead, this paper aims at investigating the regularization methods that can straightforwardly be integrated with various GANs; thereby, by applying appropriate regularization methods presented in this paper, it becomes possible to use GANs for diverse neural network architectures in many research domains. For this reason, we exclude the regularization methods for specific variants of GANs since they can hardly be used for different architectures of GAN models; we only focus on the methods that can be used universally.

In addition, in this paper, we discuss common trends in regularization methods as well as limitations of the methods and then propose future research directions. Therefore, this paper helps researchers to gain a better understanding of regularization methods in GANs, from various aspects, such as real applications of the methods and developments of novel regularization methods.

## 2. Background

### 2.1 Generative adversarial network

A conventional generative adversarial network is composed of two neural network modules, i.e., a generator and a discriminator. The generator can be interpreted as a function that produces high dimensional samples by using low dimensional feature vectors $z \in \mathbb{R}^k$ as its inputs, i.e., $G(\cdot) \coloneqq \mathbb{R}^k \to \mathbb{R}^p$, where $p$ denotes the dimension of a real sample; thus, the target of the training of the generator is to find the latent feature vectors constituting a dataset, and to allocate the features to its inputs. To achieve this goal, GANs uses an adversarial learning process with the discriminator.

The discriminator is designed to distinguish between real samples and fake samples produced by the generator. Therefore, the discriminator corresponds to a function of which dimension of the output is one, i.e., $D(\cdot) \coloneqq \mathbb{R}^p \to \mathbb{R}$. The initial GAN introduces the sigmoid function as well as the cross-entropy function for the output of the discriminator in order that the output represents the probability that the sample is real or generated.

The training of the generator and the discriminator is performed in a competitive manner, in which the generator and the discriminator play a game. In the beginning, a randomly initialized generator produces noise-like samples, which generally consist of random values. Therefore, the discriminator can be easily trained to distinguish between these fake samples and real samples. The training of the ordinary discriminator is performed by the following loss function:

$$\mathcal{L}_D \coloneqq \mathbb{E}\left[\log\left(D(G(z))\right)\right] + \mathbb{E}\left[\log(1 - D(x))\right]. \tag{1}$$

Then, the generator can learn from such learning of the discriminator, by targeting to deceive the discriminator; gradients can be backpropagated from the output of the discriminator since all components from the input of the generator to the output of the discriminator are connected through neural network structures. The training loss of the generator in the initial GANs can be represented as follows:

$$\mathcal{L}_G \coloneqq -\mathbb{E}\left[\log\left(D(G(z))\right)\right]. \tag{2}$$

By such an evolving and adversarial learning between the generator and the discriminator, latent features of a dataset can be allocated to the input variables of the generator (Figure 1(A)). After adequate iterations of such learning, the generator can produce synthetic but realistic samples by applying the Monte Carlo method to the input $Z$ (Figure 1(B)).

However, it has been known that the GAN training is not straightforward since the target of the generator is produced by the other neural network structure, i.e., the discriminator, during the training. Therefore, the target constantly fluctuates, thereby making the convergence of GANs challenging. While the convergence of GANs has been proved, an optimal discriminator is required for the convergence [20]; however, in practice, the discriminator is not optimal and not fixed during the GAN training. Thus, due to such a discrepancy of the discriminator between the theoretical proof and the actual training, there have been numerous studies to address this discrepancy.

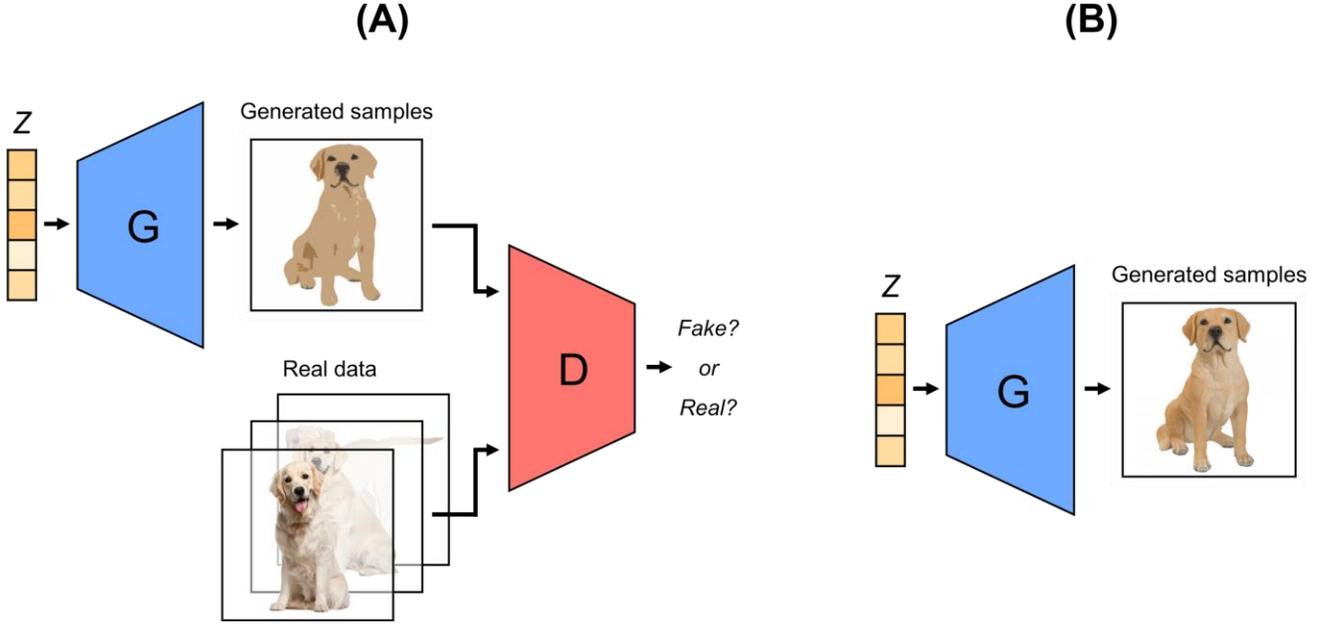

**Figure 1. The framework of GANs. (A) The training process; (B) Sample generation after the training.** G represents the generator; D represents the discriminator. The Monte Carlo method is used over the latent vector Z.

## 2.2 The mode collapse and Lipschitz continuity

The diverged GANs do not properly generate realistic samples; instead, the generator commonly produces few identical samples regardless of its inputs. Such a typical problem in the GAN training is called the mode collapse problem, which signifies that the generator only learns the highest frequency among the sample space, in order to induce the discriminator not to recognize generated samples.

Also, it has been known that the mode collapse problem is caused by the failure of the stabilization of the discriminator [2]. During the training of GANs, gradients for the training of the discriminator can explode, which makes the discriminator suddenly diverge. The diverged discriminator generally fails to cover the whole sample space; instead, it only handles several typical samples. The generator learns from such a discriminator, and therefore, the diversity of generated samples becomes significantly low.

To ensure the stabilization of GANs, it is critical to enforce the discriminator to be Lipschitz continuous (Figure 2), which can be represented as follows:

$$|D(x_1) - D(x_2)| \leq K|x_1 - x_2|, \quad (3)$$

where $K \geq 0$ is a real constant called Lipschitz constant, and $\forall x_1, x_2 \in \mathbb{R}^p$. Therefore, Lipschitz continuity implies that absolute values of the gradients of the discriminator with respect to the sample space must be in a certain range, i.e., $K$, since (3) can be rearranged by

$$\frac{|D(x_1) - D(x_2)|}{|x_1 - x_2|} \leq K, \quad (4)$$

which indicates such a characteristic.

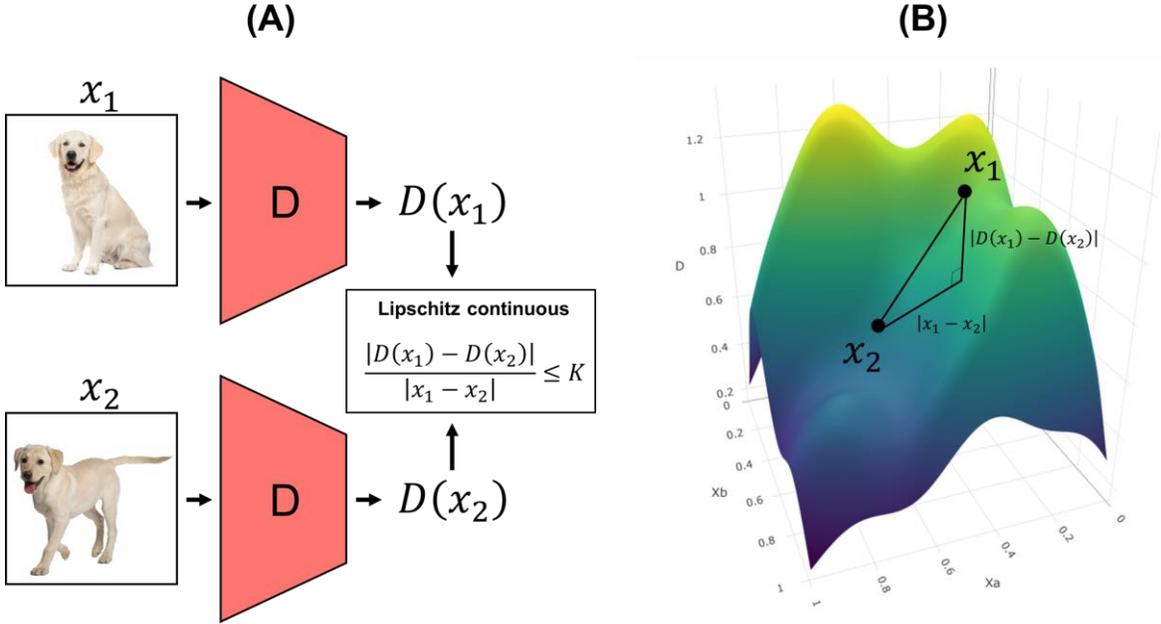

**Figure 2. Graphical representations of Lipschitz continuity. (A) Lipschitz continuity with respect to neural network; (B) Lipschitz continuity with respect to a sample space.** In (B), a surface of the outputs of the discriminator is illustrated with regard to a two-dimensional sample space with variables, Xa and Xb. Satisfying Lipschitz continuity corresponds to smoothing the surface in order to make the gradients sufficiently small.

## 3. Regularization Methods for GANs

### 3.1 An overview of regularization methods for GANs

In this paper, variants of GANs that handle the mode collapse problem and regularizations are addressed. In recent years, GANs have been widely employed for many practical applications, such as image generation [8], text generation [17], object detection [35], and image super-resolution [32], by modifications of the architectures of GANs to be adapted for specific data types and different problems. Despite the success in several applications, the notorious behavior of the GAN training still remains to be solved. Therefore, in this paper, we aim to review the regularization methods to solve this problem.

Specifically, among the regularization methods, we narrow the research scope down to universal methods that do not require special and additional neural network architectures. We only address regularization methods that are independently used irrespective of the structures. The regularization methods using modifications of neural network architectures, which we call architectural methods in this paper, are not covered in this paper. Since the structures of GANs are commonly modified for specific data types and different tasks in many real applications, it is unclear whether the architectural methods can also be applied to these modifications, while the other regularization methods can be used regardless of neural network structures in general. For example, while there are several architectural methods using an encoder structure [10,58,66,69] or additional layer attached to ordinary GANs [37,75], it is not explicit these methods can be applied to other GAN structures.

Also, it is difficult to define whether a certain architectural method corresponds to a regularization method. Numerous variants of GANs introduces novel architectures, which are commonly used for better learning of a certain sample space. However, it is hard to describe that such improvements are due to certain regularization effects since there is little evidence that the modifications in architectures and regularizations are related. Although several studies have employed auto-encoder structures for the discriminator [5,84], and demonstrated superior performance to generate synthetic images, it is also unclear that such excellent performance results from a certain regularization effect. Thus, architectural methods are excluded from the research scope due to their nonuniversality and uncertain regularization effects.

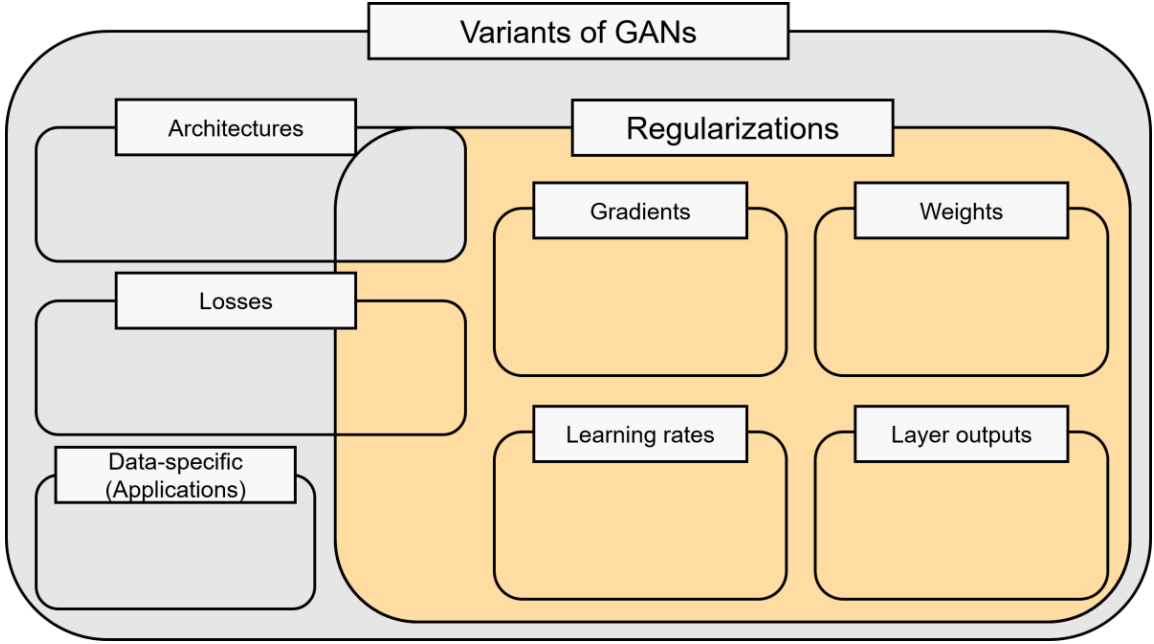

**Figure 3. Variants of GANs and the scope of this study.** The research scope is highlighted in orange. This study addresses regularization methods that are universally applied to various GANs, regardless of their architectures. Thus, regularization methods using special architectures are excluded in this study, due to their nonuniversality.

In this paper, we categorize existing regularization methods into five groups, according to their operating principles: 1) Gradient penalties; 2) Weight normalizations; 3) Imbalanced training; 4) Normalizations of outputs of layers; 5) Modified losses and targets of GANs. Then, in each group, we analyze the approach and present the most recent methods. The scope of the study is illustrated in Figure 3. We describe each group and corresponding methods in the following sections.

## 3.2 Gradient penalty methods

The gradient penalty methods are direct ways to address Lipschitz continuity problem since the gradients with respect to a sample space must be in a certain range in order to make the discriminator Lipschitz continuous, as described in Section 2.2; thus, the gradient penalty methods are the most conventional methods to stabilize the GAN training. It has been widely studied to aim at constraining the discriminator to satisfy ( 4 ) with a certain $K$. Generally, to train the discriminator, the gradient penalty methods use auxiliary losses for in addition to the ordinary loss of the generator, which can be represented as follows:

$$\mathcal{L}_{D+GP} = \mathcal{L}_D + \lambda \mathcal{L}_{GP}(\nabla_{\tilde{x}} W) \qquad (5)$$

where $\mathcal{L}_D$ denotes the ordinary loss; $\mathcal{L}_{GP}$ is the auxiliary loss function of each gradient penalty method; $\lambda$ is a setting parameter that modulates the penalty; $\tilde{x}$ represents a sample obtained from a target sample space; $W$ indicates weight matrices of the discriminator. In this manner, each gradient method varies by using different $\mathcal{L}_{GP}$ and $\tilde{x}$. The differences are summarized in Table 1.

Gradient Penalty (GP) [21] initially introduces a constraint that penalizes the gradients. In GP, the discriminator aims at targeting $\|D\|_{Lip} \to 1$ where $\|D\|_{Lip}$ denotes the left term in (4). To satisfy such a condition, GP uses a $L_2$ norm as $\mathcal{L}_{GP}$ in which gradients that deviate from one are penalized.

**Table 1. Gradient penalty methods with auxiliary penalty functions and sampling algorithms.** $\mathcal{L}_{GP}$ and $\tilde{x}$ represent the penalty function and sampling algorithms. $\alpha$ is a parameter drawn from a uniform distribution, which varies in each iteration. $\|D\|_{Lip}$ stands for gradients of Lipschitz continuity.

| Reference | Year | $\mathcal{L}_{GP}$ | $\tilde{x}$ | Lipschitz continuity |
|---|---|---|---|---|
| Gulrajani et al. [21] | 2017 | $\mathbb{E}[\|\nabla_{\tilde{x}} W\|_2 - 1]$ | $(1-\alpha)x + \alpha \hat{x}$ | $\|D\|_{Lip} \to 1$ |
| Kodali et al. [30] | 2017 | $\mathbb{E}[\|\nabla_{\tilde{x}} W\|_2 - 1]$ | $x + \epsilon$ | $\|D\|_{Lip} \to 1$ |
| Petzka et al. [53] | 2018 | $\max\{\|\nabla_{\tilde{x}} W\|_2 - 1, 0\}$ | $(1-\alpha)x + \alpha \hat{x}$ | $\|D\|_{Lip} \leq 1$ |
| Mescheder et al. [46] | 2018 | $\mathbb{E}[\|\nabla_{\tilde{x}} W\|_2]$ | $x$ or $\hat{x}$ | $\|D\|_{Lip} \to 0$ |
| Zhou et al. [86] | 2019 | $\max\{\|\nabla_{\tilde{x}} W\|_2\}$ | $(1-\alpha)x + \alpha \hat{x}$ | $\|D\|_{Lip} \to 0$ |
| Thanh-Tung et al. [64] | 2019 | $\mathbb{E}[\|\nabla_{\tilde{x}} W\|_2]$ | $(1-\alpha)x + \alpha \hat{x}$ | $\|D\|_{Lip} \to 0$ |

Also, in GP, a process to obtain $\tilde{x}$ is introduced; an interpolation between a real sample and a generated sample is used, which can be obtained as follows:

$$\tilde{x} = (1-\alpha)x + \alpha\hat{x}, \qquad (6)$$

where $\alpha$ indicates a parameter sampled from $Unif(0,1)$ in each training iteration; $x$ denotes a real sample; $\hat{x}$ represents a generated sample with the generator.

Other gradient penalty methods correspond to variants of GP in general, where similar approaches are employed for $\mathcal{L}_{GP}$. Petzka et al. [53] argued that GP does not directly enforce Lipschitz continuity since $\|D\|_{Lip}$ converges to around one, and gradients having small values of $\|D\|_{Lip}$ are trained to be one. Thus, they proposed a maximum threshold for the initial GP, called Lipschitz Penalty (LP), in order to handle this problem. Consequently, in LP, the gradients can be constrained to be under one.

Zhou et al. [86] introduce Max Gradient Penalty (MaxGP), which uses the maximum value of gradients as the penalty. From the claim that the maximum of the gradients is equivalent to Lipschitz constant $K$, proposed in Adler and Lunz [1], MaxGP aims at penalizing the maximum value, i.e., $\mathcal{L}_{GP} \coloneqq \max\{\|\nabla_{\tilde{x}} W\|_2\}$. Therefore, since the gradients are directly penalized without a certain threshold, they converge to zero, which is distinct from LP.

Recently, Thanh-Tung et al. [64] proposed another gradient penalty method using the same convergence with MaxGP, called 0-GP. In contrast to MaxGP, 0-GP uses an average value of all weights, making the weights strictly regularized to satisfy the convergence. However, in the study, although 0-GP was compared to several similar methods using the same convergence but different sampling algorithms, MaxGP and 0-GP were not compared; thus, it is difficult to conclude that such a strict regularization is effective.

The aforementioned methods commonly use the same sampling process as that of GP, in which the interpolated value between a real sample and a generated sample is employed. In contrast to these methods, there are several methods that introduce different sampling process. For instance, Kodali et al. [30] introduced $x + \epsilon$ as $\tilde{x}$, which uses small noises added to real samples. Similarly, Mescheder et al. [46] employed both real samples and fake samples as $\tilde{x}$. They claimed that it is not necessary to satisfy Lipschitz continuity in the whole sample space. Since the discriminator is trained with real samples and generated samples, it is sufficient to satisfy Lipschitz continuity around the training space, i.e., $x + \epsilon$ and $\hat{x} + \epsilon$. Therefore, the sampling algorithm used in GP and the other methods restrict the training of the discriminator in an excessive manner because the algorithm tries to limit the whole interpolated spaces between real samples and generated samples. Such a claim

**Table 2. Weight penalty and normalization methods.** $\mathcal{L}_W$ represents the penalty function. $\widetilde{W}$ denotes the normalized weight and algorithm to obtain the weight. $\sigma$ stands for the largest singular value of a matrix.

| Reference | Year | Method | $\mathcal{L}_W$ | $\widetilde{W}$ |
| --- | --- | --- | --- | --- |
| Arjovsky et al. [2] | 2017 | Normalization | - | $\widetilde{W} \coloneqq clip(W, [-0.01, 0.01])$ |
| Brock et al. [7] | 2017 | Penalty | $\|WW^T - I\|_F^2$ | - |
| Miyato et al. [47] | 2018 | Normalization | - | $\widetilde{W} \coloneqq W/\sigma(W)$ |
| Zhou et al. [85] | 2018 | Penalty | $\|W\|_\infty$ | - |
| Brock et al. [8] | 2019 | Penalty | $\|WW^T \odot (\mathbf{1} - I)\|_F^2$ | - |
| Kurach et al. [31] | 2019 | Penalty | $\|W\|_2$ | - |
| Liu et al. [39] | 2019 | Normalization | - | $\widetilde{W} \coloneqq W/\sigma(W) + \nabla W/\sigma(W)$ |
| Zhang et al. [83] | 2019 | Normalization | - | $\widetilde{W} \coloneqq W/\sqrt{\|W\|_1 \|W\|_\infty}$ |

should be further investigated since the motivation is valid.

Overall, we found that the gradient penalty methods correspond to variants of ( 5 ). The methods commonly use an auxiliary loss, in addition to ordinary GAN loss, in order to penalize the gradients. Also, we also found that while the initial methods aim at making the gradients of the discriminator to be one or less than one, recent methods generally penalize the gradients without such a threshold, making the gradients converge to zero.

### 3.3 Weight penalty and normalization methods

Weight penalization is a standard regularization method in many statistical and machine learning models, including artificial neural network. Conventional statistical models using weight parameters have introduced the penalization methods to reduce overfitting. Ridge regression and Least Absolute Shrinkage and Selection Operator (LASSO) employ such penalization methods by adopting additional $L_2$ and $L_1$ norms, respectively, to their optimization functions. In a similar manner, there have been several studies to use this approach in artificial neural network models.

In GAN models, weight penalization methods have been extensively used to prevent the mode collapse problem. Generally, the weight penalization methods for GAN can be classified into two types: weight penalty and weight normalization. Weight penalty methods employ additional losses for the target function of GAN, which is significantly similar to the approaches of the ridge regression and LASSO. Therefore, the loss function with the weight penalty method is the same form as **( 5 )**, where a norm of weight matrices is used instead of $\mathcal{L}_{GP}$. In contrast, weight normalization methods constantly update the weight matrices in GAN, by a specific training algorithm during the GAN training. The recent weight penalty and normalization methods are listed in Table 2.

First, for a weight penalty method, there have been attempts to use the conventional $L_2$ for the GAN training. Kurach et al. [31] evaluated $L_2$ norm as a weight penalty method; as a result, it was demonstrated that $L_2$ norm decreases the performance instead. Such a result also signifies that special methods are required for the training of GAN since the most conventional method for neural network models is not valid.

In a similar manner to $L_2$ norm, other norms were assessed in Zhou et al. [85], where they further evaluated $L_1$ and $L_\infty$ norms. As a result, these norms showed regularization effects during the GAN training, and computational time reduces compared to GP. However, as they mentioned regarding the results, $L_\infty$ norm did not outperform GP with respect to sample generation performance.

Brock et al. [7] also investigated this problem and proposed Orthogonal Regularization (OR) as a weight penalty. By using the characteristic of matrix multiplication that an orthogonal weight matrix does not alter the norm, OR leads the weights to be orthogonal, with

$$\mathcal{L}_W := \sum \|WW^T - I\|_F^2, \qquad (7)$$

in which the non-diagonal elements of $WW^T$ converge to zero. However, such a method also restricts $L_2$ norm of the matrix, thereby making the GAN training challenging, as claimed in [47]. Thus, Brock et al. [8] proposed a novel OR method that does not directly limit $L_2$ norm, which can be represented as

$$\mathcal{L}_W := \sum \|WW^T \odot (\mathbf{1} - I)\|_F^2, \qquad (8)$$

where $\odot$ denotes Hadamard product; and $\mathbf{1}$ represents a matrix of which all elements are one.

Distinct from weight penalty methods, weight normalization methods do not use additional loss. Instead, they constantly update the weights during the training process, or compute gradients with respect to the normalized weights then backpropagate the gradients.

Spectral Normalization (SN) [47] is one of the most conventional weight normalization methods, which introduces the spectral norm of weight matrices for the GAN training. The spectral norm is a type of matrix norm, which is equivalent to $L_2$ norm of a vector. In addition, it is known that the spectral norm corresponds to the largest singular vector. Using the claim that the largest singular vector is related to Lipschitz constant, as verified in [77], the backpropagation of SN is performed with normalized weight matrices, as follows:

$$W_{k+1} := W_k - \beta \nabla_W D_{SN(W)}, \qquad (9)$$

where $k$ is an index of iteration; $\beta$ is a learning rate; $D_{SN(W)}$ indicates a discriminator with spectrally normalized weight matrices. Such a process signifies that the spectral norm is employed only for gradient calculation, and weight matrices themselves are not changed in SN.

Furthermore, SN uses the power iteration method to calculate the spectral norm, while a conventional optimization method can hardly be used since gradients through the optimization method cannot be computed. In the power iteration method, the spectral norm is approximated with multiplications of vectors and a weight matrix, thereby, making it possible to compute gradients:

$$SN(W) \cong \frac{W}{u^T W v}, \qquad (10)$$

where $W \in \mathbb{R}^{n \times m}$; $u \in \mathbb{R}^n$; $v \in \mathbb{R}^m$; and $u^T W v$ signifies the approximated spectral norm. The $u$ and $v$ are randomly initialized at first, then updated through the power iteration method.

A variant of SN has been introduced in Liu et al. [39], in which the gradients of weight parameters are additionally spectrally penalized with the original SN. They investigated the concept of spectral collapse in the study, and claimed that such a method, called Spectral Regularization (SR), can tackle the spectral collapse. Evaluated with image datasets, SR demonstrated superior regularization performance, compared to SN.

Similarly, Zhang et al. [83] further explored the use of the spectral norm for the GAN training, and proposed Spectral Bounding (SB) method. SB aims to restrict the upper bound of the spectral norm, which can be computed by

$$\mathcal{L}_{SB} := \sqrt{\|W\|_1 \|W\|_\infty}. \qquad (11)$$

Furthermore, in the study, it was reported that SB outperformed SN in the experiments with two image datasets.

In addition, weight clipping method, which can be interpreted as a weight normalization method, is used to train initial Wasserstein GAN (WGAN) [2], while WGAN has been extensively studied thereafter. The weight clipping method encourages the discriminator to satisfy Lipschitz continuity by regulating values of weight matrices to be in a certain small range. In the

**Table 3. Algorithms and applications for imbalanced training in GANs.**

| Reference | Year | Approach | Description |
|---|---|---|---|
| Goodfellow et al. [20] | 2014 | Multiple updates | They proposed a multiple update algorithm with a parameter $k$ for the discriminator; however, $k$ was set to one in their experiments. |
| Arjovsky et al. [2] | 2017 | Multiple updates | $k$ was set to five. |
| Heusel et al. [22] | 2017 | Learning rate | They advocated using imbalanced learning rates (TTUR). |
| Brock et al. [8] | 2019 | Multiple updates & Learning rate | Both algorithms were simultaneously used, in which double updates and a double learning rate were adopted for a discriminator |

---

**Algorithm 1**: Multiple update algorithm [20]

**Input:** A dataset ($P_x$); a noise distribution ($P_z$).
**Model:** A generator $G$; a discriminator $D$.
1:  **Sample** a noise batch $z$ from $P_z$
2:  **Update** $G$ with $G \leftarrow G + \beta \nabla D(G(z))$
3:  **for** $k$ times **do:**
4:      **Sample** a noise batch $z$ from $P_z$
5:      $\hat{x} = G(z)$
6:      **Sample** a sample batch $x$ from $P_x$
7:      **Update** $D$ with $D \leftarrow D + \beta \nabla D(x) - \beta \nabla D(\hat{x})$, where $x \sim P_x$.
8:  **endfor**

---

study, they clipped the weight values to a fixed range of $[-0.01, 0.01]$. However, such an approach significantly hinders the GAN training because the weights cannot be further trained beyond the values, which becomes the motivation of GP.

To sum up, the weight normalization methods have been employed for satisfying Lipschitz continuity while the weight penalty methods have been introduced for orthogonality of weight matrices or general stability of the GAN training. Also, the weight normalization methods have been reported that they commonly enhance sample generation performance of GANs [2,47,83], while a weight penalty method demonstrated inferior performance than ordinary GAN without the method [31], and a few weight penalty methods can be used only with the other GAN training methods, such as truncation trick [8] and orthogonal initialization [7,60]. Since Lipschitz continuity is an essential condition of the stability of the GAN training, as described in the previous section, it has been demonstrated that the methods which properly tackle this problem show excellent performance in general.

## 3.4 Imbalanced training

In the initial GAN, Goodfellow et al. [20] proved that ideal GAN training between the discriminator and generator reaches an equilibrium under the condition that an optimal discriminator is given. However, actual GAN training generally does not meet the condition. Thus, to make the discriminator be close to the ideal, Goodfellow et al. [20] proposed an algorithm using multiple updates for the discriminator, which is shown in Algorithm 1. Although it has been demonstrated that a single update is sufficient in their experiments, such a multiple update algorithm has been considered as an appropriate option to train GAN.

Table 4. Normalization methods for outputs of a layer in GANs.

| Method | Proposed for neural networks | Proposed and evaluated with GANs |
|---|---|---|
| Batch normalization | Ioffe and Szegedy [26] | Kurach et al. [31]; Miyato et al. [47] |
| Layer normalization | Ba et al. [4] | Kurach et al. [31]; Miyato et al. [47] |
| Weight normalization | Salimans and Kingma [59] | Miyato et al. [47] |
| Instance normalization | Ulyanov et al. [68] | Karras et al. [28] |
| Group normalization | Wu and He [74] | - |
| Conditional batch normalization | Dumoulin et al. [16] | Miyato et al. [47]; Zhang et al. [80] |
| Self-modulation | - | Chen et al. [11] |

For example, in WGAN [2], the discriminator uses the multiple update algorithm because the weight clipping method is applied, thereby hindering the training, where the discriminator of WGAN was trained 5 times per generator training. Likewise, in other various GANs [8,21,47], such an approach with imbalanced training has been conventionally used.

Heusel et al. [22] investigated this problem from a different perspective and argued that distinct learning rates for the discriminator and generator can solve this problem. They claimed a local Nash equilibrium can be obtained even if the learning rates of the discriminator and generator are different from each other, and proved this claim in depth. Hence, instead of using the multiple update algorithm for the discriminator, a higher learning rate than that of the generator can reduce computational time. They named this algorithm as Two Time-scale Update Rule (TTUR).

Overall, because the balanced training between the discriminator and generator is essential, there have been two simple approaches enforcing imbalanced training to handle this issue: Multiple updates and imbalanced learning rates. The descriptions of these approaches are shown in Table 3. The two approaches are intuitive and straightforward since the methods directly train the discriminator severer than the generator to maintain the balance. In addition, it has been verified that such simple methods operate properly in many studies, even if the methods are simultaneously used [8]. However, the number of multiple updates and values of different learning rates vary according to GAN architecture, experiments, and datasets, which remains to be further studied to determine how to select these parameters, while existing studies commonly used trial-and-error methods.

## 3.5 Normalization on outputs of a layer

After the development of Batch Normalization (BN) [26,42], which brings significant progress in the training of deep neural networks, such normalization methods to restrict the outputs of each neural network layer have now become essential to train neural network models [36,40,72]. Also, variants of BN, such as Layer Normalization (LN) [4], Weight Normalization (WN) [59], Instance Normalization (IN) [68], and Group Normalization (GN) [74], have been introduced for the same purpose. The normalization methods are summarized in Table 4. Generally, normalization methods can be performed as follows:

$$h_N = \gamma \odot \frac{h - \mu}{\sigma} + b, \quad (12)$$

where $h$ denotes outputs of a layer; $\gamma$ and $b$ are learnable scale and shift parameters; $\mu$ and $\sigma$ are the average and standard deviation of $h$, respectively. The difference between the normalization methods is in obtaining $\mu$ and $\sigma$, in which, for instance, they are calculated within the mini-batch in BN, or within the nodes in a layer in LN.

There have been several approaches to adopt the normalization methods to the discriminator in GANs as well [15,56]. BN, which is the most conventional normalization method in general neural networks, was evaluated in Lucic et al. [41]. However, as a result, it was demonstrated that the conventional BN is not valid for GANs, in which the performance decreased when BN was used.

In contrast, other conventional methods that are used in typical neural networks, including LN and WN, were investigated in WGAN with GP [21] and further evaluated in Miyato et al. [47] and Kurach et al. [31]. The methods were assessed for the discriminator in GANs and showed superior performance than ordinary GANs. Specifically, in both studies, it has been verified that LN enhances sample generation performance in general, regardless of parameter settings.

However, for the generator, BN has been commonly used and successfully demonstrated its effectiveness [47]. Furthermore, conditional BN (cBN) as well as Adaptive IN (AdaIN) [14,16,19,25] have become the dominant methods to provide conditional information for the discriminator. Consequently, in recent GANs, it is general to use BN only for the generator and not for the discriminator [8,28].

As another variant of BN, Chen et al. [11] proposed Self-Modulation (Self-Mod) method for GANs. While ordinary BN uses learnable scale and shift parameters, in Self-Mod, they modified these parameters to be input-dependent. Therefore, the learnable parameters take $z$ as their inputs; then the dependencies are trained through neural network structures, which can be represented as

$$h_N = \gamma(z) \odot \frac{h - \mu}{\sigma} + b(z), \quad (13)$$

$$\gamma(z) := W_\gamma^{(1)} \cdot ReLU\big(W_\gamma^{(2)} z + \beta_\gamma\big), \quad (14)$$

$$b(z) := W_b^{(1)} \cdot ReLU\big(W_b^{(2)} z + \beta_b\big), \quad (15)$$

where $W$ denotes a weight matrix; $\beta$ represents a bias parameter of the neural network structures; the activation function of the neural network structures is set to Rectified Linear Unit (ReLU).

## 3.6 Regularization with modified loss functions

Most variants of GANs correspond to modifications of loss functions [2,43,51,63]. Since initial GANs uses the cross-entropy loss function, and suffers from the mode collapse problem, to handle this problem, the variants of GANs introduces different loss functions derived from various mathematical theories. However, it is difficult to say that all these variants and loss functions are for the regularization of GANs because they are generally motivated from better learning of sample distributions. In this study, the loss functions that directly aim at regularization are addressed, where Lipschitz continuity problem is handled.

The loss functions for GANs to handle Lipschitz continuity problem were initially explored by Arjovsky et al. [2] and Qi [55] about the same time. Arjovsky et al. [2] introduced Earth-Mover (EM) distance to solve this problem and named the method as WGAN. In contrast, while the same problem was addressed, Qi [55] proposed a GAN model using a feature-wise distance, called Loss-Sensitive GAN (LSGAN), where the distance is obtained through pre-trained networks. Also, it is verified that WGAN is a special case of Generalized LSGAN (GLSGAN).

In WGAN, the discriminator and generator are trained with the following loss functions:

$$\mathcal{L}_D := \mathbb{E}\big[D\big(G(z)\big)\big] - \mathbb{E}[D(x)], \quad (16)$$

$$\mathcal{L}_G := -\mathbb{E}\big[D\big(G(z)\big)\big], \quad (17)$$

under the condition that $\|D\|_{Lip} \leq K$.

While the same loss function for the generator, i.e., ( 17 ), is used in LSGAN, a distinct loss function for the discriminator is introduced, which can be calculated as follows:

$$\mathcal{L}_D := -\omega \cdot \left(\Delta(X, G(z)) + \mathbb{E}[D(x)] - \mathbb{E}[D(G(z))]\right) - \mathbb{E}[D(x)], \tag{18}$$

where $\omega$ denotes a positive balancing parameter; $\Delta(a, b)$ is a feature-wise distance of which features are obtained from pre-trained networks, such as Inception network [62].

Recently, a more improved loss function for the discriminator, called hinge loss, was proposed and employed in various GANs [38,65]. Also, although the loss function for the generator is the same as WGAN and LSGAN, the discriminator is more regularized with the hinge loss, which can be computed as follows:

$$\mathcal{L}_D := \mathbb{E}\left[\max\left(0, 1 + D(G(z))\right)\right] + \mathbb{E}[\max(0, 1 - D(x))], \tag{19}$$

which can be interpreted as a restricted Wasserstein distance where only less trained samples are optimized in a discriminator with output thresholds of $1$ and $-1$.

In a similar manner, Ni et al. [49] proposed a regulated loss for the discriminator, in which generated samples are gradually trained. While the same loss as that of WGAN is used for the generator, the discriminator does not learn by generated samples at first, then gradually learns them, which can be represented as

$$\mathcal{L}_D := k_t \cdot \mathbb{E}[D(G(z))] - \mathbb{E}[D(x)], \tag{20}$$

where $k_{t+1} := k_t + \lambda \cdot (\mathcal{L}_{D(x)} + \mathcal{L}_G)$; $\lambda$ is a regularization parameter; and $\mathcal{L}_{D(x)}$ denotes $\mathbb{E}[D(x)]$.

## 3.7 Other regularization methods

There are a few other regularization methods that do not correspond to the above-mentioned categories. Each method uses its own approach instead of the presented approaches in the previous sections. In this section, these unique methods and approaches are addressed.

It has been explored to regulate the consistency of generated samples and $D(x)$ with respect to inputs of the generator and the discriminator, respectively. Gan et al. [18] proposed a method to smooth the differences in generated samples with regard to its inputs, i.e., $z$. The motivation of the method is straightforward, in which two generated samples, i.e., $G(z_1)$ and $G(z_2)$, should be similar if the difference between $z_1$ and $z_2$ is small. Therefore, to maintain such a consistency, they used the following penalized loss for the generator:

$$\mathcal{L}_G := -\mathbb{E}[D(G(z))] + \alpha \cdot [(\min(k, \delta_{min}) - \delta_{min})^2 + (\max(k, \delta_{max}) - \delta_{max})^2], \tag{21}$$

where $k := \|G(z + \varepsilon) - G(z)\|_2 / \|\varepsilon\|_2$; $\alpha$ is a balancing parameter; $\delta_{min}$ and $\delta_{max}$ are thresholds to maintain $k$ to be in a certain range. Similarly, Zhang et al. [81] proposed a regularization method for the consistency of the discriminator with respect to data augmentation. They argued that the outputs of the discriminator of two samples, i.e., $D(x_1)$ and $D(x_2)$, should be similar if features of $x_1$ and $x_2$ are almost same. Therefore, they set $x_2 := T(x_1)$, where $T(\cdot)$ is a data augmentation method, such as cropping and rotating. Then, a penalization loss, called Consistency Regularization (CR), is added to the hinge loss, i.e., $\mathcal{L}_{D+CR} := \mathcal{L}_D + \alpha \cdot \mathcal{L}_{CR}$, where $\mathcal{L}_D$ is the same as (19); and

$$\mathcal{L}_{CR} := \|D(x) - D(T(x))\|_2^2. \tag{22}$$

Another regularization method using dropout [61] was proposed in Wei et al. [73], where the consistency of dropout is maintained. Dropout layers are used in the discriminator; then, discriminator outputs of two samples, i.e., $x_1$ and $x_2$, should be similar to satisfy Lipschitz continuity. Consequently, such a characteristic is implemented by dropout with a same input. Hence, Consistency Term (CT) that was proposed in the study can be calculated as follows:

**Table 5. State-of-the-art GANs and regularization methods used in the studies.** Recent studies that have been cited more than 500 times according to Google Scholar were investigated. MU stands for multiple update algorithm; HL stands for the hinge loss. A semicolon indicates that both methods were simultaneously used. A forward slash indicates two methods were selectively used depending on experiment. The other abbreviations can be found in each section of this paper.

| Reference | Year | Task | Gradient penalty | Weight regularization | Imbalanced training | Layer output normalization | Regularized loss |
| --- | --- | --- | --- | --- | --- | --- | --- |
| Wang et al. [71] | 2018 | Style transfer | None | None | None | IN | None |
| Hoffman et al. [23] | 2018 | Style transfer | None | None | None | BN | None |
| Karras et al. [27] | 2018 | Image generation | GP | None | MU | BN; LN | WGAN |
| Choi et al. [12] | 2018 | Style transfer | GP | None | MU | IN | WGAN |
| Miyato et al. [47] | 2018 | Image generation | GP/None | SN | MU | BN/cBN | HL |
| Zhang et al. [80] | 2019 | Image generation | None | SN | TTUR | cBN | HL |
| Brock et al. [8] | 2019 | Image generation | GP | SN; WN | MU; TTUR | cBN | HL |
| Karras et al. [28] | 2019 | Image generation | GP | None | None | AdaIN | WGAN |

$$\mathcal{L}_{CT} := \sum \lambda_i \cdot \left| D_1^{(i)}(x) - D_2^{(i)}(x) \right|, \qquad (23)$$

where $D_k^{(i)}$ denotes the output of $i^{th}$ layer of the discriminator with a random dropout represented as $k$; $\lambda_i$ is a weight parameter of the layer. $\lambda_i$ was set to 1 and 0.2 for the last layer and the penultimate layer, respectively. This CT is used in addition to the ordinary discriminator loss in a similar manner to other studies.

## 4. Regularization Methods in State-of-the-art GANs

In this section, regularization methods used in state-of-the-art GANs are investigated. Specifically, we selected highly cited GAN studies and investigated regularization methods used in the studies, in order to explore popular regularization methods; we selected the papers that have been published within the last three years, i.e., the years of 2018-2020, and cited more than 500 times according to Google Scholar [88].

The selected studies and analysis results are shown in Table 5. The studies can be categorized into two classes according to their tasks, which are image generation and style transfer. Image generation is the original purpose of initial GANs [20], which use noise vector as inputs of the generator and produce synthetic images from the noises. Style transfer using GANs has been studied widely after the development of Cycle-consistent GAN (CycleGAN) [87], which takes unpaired samples as its inputs, then trains with the GAN loss as well as cycle-consistent losses obtained through two generators in CycleGAN.

As a result, the first gradient penalty method, i.e., GP, is the dominant method for penalizing gradients, while a variety of other gradient penalty methods have been recently proposed as described in Table 1. In addition, for weight regularization in very recent models, SN has been used widely since its development [47]. In contrast, the multiple update algorithm and TTUR are both used in recent state-of-the-art GANs. Specifically, Brock et al. [8] performed a large scale study for imbalanced training, and reported that the best performance was obtained when both methods were used simultaneously. All models have employed layer output normalization methods; however, recent models used the method only for the generator. While the initial GAN loss, which is represented as None in Table 5, WGAN loss, and hinge loss are utilized in the selected GANs, the hinge loss is commonly used in the latest models.

# 5. Limitations and Future Directions

Although numerous studies have explored regularization in GANs with various aspects, there still remain several limitations to be further investigated. In this section, we discuss such limitations and propose future research directions to tackle these limitations.

While GP is the dominant method for penalizing gradients, there are several limitations: First, GP can hardly satisfy Lipschitz continuity in a direct manner because the gradients become around one, while they must be under a certain threshold to satisfy Lipschitz continuity. Such an issue also makes the discriminator difficult to converge because a discriminator having small gradients below one is enforced to have large gradients near one by GP, making the discriminator can hardly converge. Second, GP uses an interpolated value between real samples and generated samples; however, the interpolated value cannot represent the whole sample space nor the sample space between real samples and fake samples. Specifically, the sample space between real and fake should be obtained with regard to features instead of pixels. Third, in the first place, it is sufficient that the discriminator satisfies Lipschitz continuity only near the real samples and fake samples, since the discriminator is trained with these samples, and not the interpolated values nor the whole sample space.

As recent studies using $\|D\|_{Lip} \to 0$, future work should investigate gradient penalty methods that make the gradients under a certain threshold. By the Lagrange multiplier method [57], the gradients are constrained to be a certain range even if the auxiliary loss for the penalty converges to zero. In such a case, the penalty parameter $\lambda$ determines Lipschitz constant. This relationship between the Lagrange multiplier method and the penalty method using $\|D\|_{Lip} \to 0$ should be further investigated as well.

Also, as previously described, it is sufficient to satisfy Lipschitz continuity near the real samples and generated samples. In other words, the discriminator should be locally Lipschitz continuous instead of strictly satisfying it. While most of the existing methods use pixel-wise, interpolated values, it is more natural to simultaneously penalize the gradients with respect to both $x + \epsilon$ and $G(z) + \epsilon$.

Furthermore, while GP aims at 1-Lipschitz continuous of the discriminator, the Lipschitz constant of one is an arbitrary parameter. Lipschitz constant can be set to any number, and we can conjecture that Lipschitz constant is highly related to learning rates and values of weight parameters. For instance, Karras et al. [27] evaluated various Lipschitz constant with their model, and reported that they obtain a significantly better result when Lipschitz constant is set to 750. Such a study regarding optimal Lipschitz constants should be conducted as well.

In a similar manner, SN also aims at 1-Lipschitz continuous of the discriminator, by dividing weight parameters with the maximum singular value, i.e., $\widetilde{W} \coloneqq W/\sigma(W)$. However, such an operation can be relaxed to satisfy Lipschitz continuity with another constant, by

$$\widetilde{W} \coloneqq K \cdot W/\sigma(W), \qquad (24)$$

because 1-Lipschitz continuity in SN is also arbitrarily set.

The GAN training in recent studies is generally conducted with the multiple update algorithm and TTUR using Adam optimization [29], which is a conventional method for the training of neural networks. While such a training method has demonstrated fine performance, specific optimization methods for GANs are required due to the unique training process of GANs, of which generator targets to the discriminator, which is another neural network structure. Several existing methods have explored this problem [45,48,50], and proposed optimization methods using the gradients of its counterpart. These approaches should be further investigated.

Recently, Chu et al. [13] argued that a smooth activation function is required for the discriminator to stabilize the GAN training. While existing studies have focused only on the approaches presented in this study, such as in terms of weight parameters and Lipschitz continuity, such an aspect based on the smoothness of activation functions should be investigated further.

# 6. Conclusion

In this paper, we reviewed and classified regularization methods for GANs. The existing methods were categorized into five classes according to their operation principles; then we analyzed each group. While numerous methods have been proposed in recent years, we found that there still remain several limitations in the method. We also proposed future research directions to tackle these problems. We believe that this study can help researchers to select appropriate regularization methods for specific neural network architectures and datasets, and to gain a better understanding of existing methods in order to develop novel regularization methods that handle the limitations in existing methods.

| | |
|---|---|
| | regularization," In: *IEEE International Conference on High Performance Computing and Communications; IEEE International Conference on Smart City; IEEE International Conference on Data Science and Systems (HPCC/SmartCity/DSS)*, Zhangjiajie, China. |
| [76] | X. Yi, E. Walia, and P. Babyn, (2019), "Generative adversarial network in medical imaging: A review," *Medical image analysis,* p. 101552. |
| [77] | Y. Yoshida and T. Miyato, (2017), "Spectral norm regularization for improving the generalizability of deep learning," *ArXiv preprint arXiv:1705.10941*. |
| [78] | J. Yu, Z. Lin, J. Yang, X. Shen, X. Lu, and T. S. Huang, (2018), "Generative image inpainting with contextual attention," In: *IEEE Conference on Computer Vision and Pattern Recognition (CVPR)*, Salt Lake City, UT. |
| [79] | H. Zhang, T. Xu, H. Li, S. Zhang, X. Wang, X. Huang, and D. N. Metaxas, (2018), "StackGAN++: Realistic image synthesis with stacked generative adversarial networks," *IEEE Transactions on Pattern Analysis and Machine Intelligence,* vol. 41, no. 8, pp. 1947-1962. |
| [80] | H. Zhang, I. Goodfellow, D. Metaxas, and A. Odena, (2019), "Self-attention generative adversarial networks," In: *International Conference on Machine Learning (ICML)*, Long Beach, CA. |
| [81] | H. Zhang, Z. Zhang, A. Odena, and H. Lee, (2020), "Consistency regularization for generative adversarial networks," In: *International Conference on Learning Representations (ICLR)*, Addis Ababa, Ethiopia. |
| [82] | Z. Zhang, Y. Xie, and L. Yang, (2018), "Photographic text-to-image synthesis with a hierarchically-nested adversarial network," In: *IEEE Conference on Computer Vision and Pattern Recognition (CVPR)*, Salt Lake City, UT. |
| [83] | Z. Zhang, Y. Zeng, L. Bai, Y. Hu, M. Wu, S. Wang, and E. R. Hancock, (2019), "Spectral bounding: strictly satisfying the 1-Lipschitz property for generative adversarial networks," *Pattern Recognition,* p. 107179. |
| [84] | J. Zhao, M. Mathieu, and Y. LeCun, (2019), "Energy-based generative adversarial networks," In: *International Conference on Learning Representations (ICLR)*, Tulon, France. |
| [85] | C. Zhou, J. Zhang, and J. Liu, (2018), "Lp-WGAN: Using Lp-norm normalization to stabilize Wasserstein generative adversarial networks," *Knowledge-Based Systems,* vol. 161, pp. 415-424. |
| [86] | Z. Zhou, J. Liang, Y. Song, L. Yu, H. Wang, W. Zhang, Y. Yu, and Z. Zhang, (2019), "Lipschitz generative adversarial nets," In: *International Conference on Machine Learning (ICML)*, Long Beach, CA. |
| [87] | J.-Y. Zhu, T. Park, P. Isola, and A. A. Efros, (2017), "Unpaired image-to-image translation using cycle-consistent adversarial networks," In: *IEEE International Conference on Computer Vision (ICCV)*, Venice, Italy. |
| [88] | Google Scholar. Available: https://scholar.google.com/ |
18